\documentstyle[preprint,aps,epsfig]{revtex}  

\title{Mixing and equilibration: Protagonists in the scene of 
nonextensive statistical mechanics}
\author{Constantino Tsallis$^a$\thanks{tsallis@cbpf.br}, 
Ernesto P. Borges$^{a,b}$ and Fulvio Baldovin$^a$ }
\address{$^a$Centro Brasileiro de Pesquisas Fisicas,
Rua Xavier Sigaud 150, 22290--180 Rio de Janeiro-RJ, Brazil\\
$^b$Escola Polit\'ecnica, Universidade Federal da Bahia, 40210--630 
Salvador-BA, Brazil\\}

\draft

\begin{document}

\maketitle

\begin{abstract}

After a brief review of the present status of nonextensive statistical
mechanics, we present a conjectural scenario where mixing (characterized by the
entropic index $q_{mix} \le 1$) and equilibration (characterized by the entropic
index $q_{eq} \ge 1$) play central and inter-related roles, and appear to
determine {\it a priori} the values of the relevant indices of the formalism.
Boltzmann-Gibbs statistical mechanics is recovered as the $q_{mix}=q_{eq}=1$
particular case. 

\end{abstract}

\bigskip

Human knowledge progresses along very many paths, and very rarely these paths follow a systematic and logical ordering. What we can presently witness about the formalism frequently referred to as {\it nonextensive statistical mechanics} (first proposed in 1988  \cite{tsallisjsp} and further implemented in \cite{curado,tsamepla}; for reviews, see \cite{tsallisbjp,tsallisspringer,tsallisdenton})
is by no means exception. Since it is the purpose of the present lines to start with a review, let us present it in an order which makes some epistemological sense, although it does not necessarily follow the chronology of the events. We will successively comment on (i) mathematical properties, which make the formalism to appear just as {\it applied mathematics}; (ii) the successful confrontations with experimental (as well as computational) results, which we believe raises the formalism to the status of {\it theoretical physics}, since it has a compromise with phenomena indeed occurring in nature; and (iii) the lines along which the calculation of the relevant entropic indices $q_{mix}$ and $q_{eq}$ appears as possible, so that the formalism becomes a {\it closed theory} (in other words, a {\it complete theory}: see \cite{einstein}), where everything can in principle be calculated a priori once the dynamics of the system is fully characterized.

\noindent
{\it {\bf Applied mathematics}}

The axiomatic starting point is the proposal of a possible generalization of Boltzmann-Gibbs (BG) statistical mechanics by postulating the following entropic form
\begin{equation}
S_q=\frac{1-\sum_{i=1}^W p_i^q}{q-1} = \langle \ln_q \frac{1}{p_i} \rangle \;\;\; (\sum_{i=1}^W p_i =1) \;,
\end{equation} 
where we are using $k_B =1$ (without loss of generality), $q \in {\cal R}$, $\langle \cdots\rangle \equiv \sum_{i=1}^W p_i (\cdots)$ and $\ln_q x \equiv \frac{x^{1-q}-1}{1-q}$ (with $\ln_1 x= \ln x$) [consistently, its inverse function is $e_q^x \equiv [1+(1-q)x]^{1/(1-q)}$, with $e_1^x = e^x$]. Clearly, for $q=1$, we recover the usual expression $S_1 =- \sum_{i=1}^W p_i \ln p_i$, from now on referred to, for simplicity, as the BG entropy. Also, if $A$ and $B$ are two probabilistically {\it independent} systems (i.e., $p_{ij}^{A+B} = p_i^A p_j^B$), $S_q$ satisfies the following {\it pseudoadditivity property}: $S_q(A+B) = S_q(A)+S_q(B)+(1-q)S_q(A)S_q(B)$,  hence $q=1,\;>1, \;<1$ respectively correspond to extensive, subextensive and superextensive cases. 

For an isolated system (microcanonical ensemble), optimization of $S_q$ yields equiprobability, i.e., $p_i = 1/W \;(\forall i)$, hence
\begin{equation} 
S_q = \ln_q W \;.
\end{equation} 

Before going on, it is important to stress that the pseudoadditivity property  mentioned above does not exclude that, for some classes of interdependency between $A$ and $B$, a special value $q^*$ could exist such that extensivity could be recovered in the sense that $S_{q^*}(A+B) = S_{q^*}(A)   +S_{q^*}(B)$ . If so, we could say that the adequacy of the entropic form enables the preservation of the traditional property of extensivity for the entropy. As mathematical illustrations of this fact, let us consider two possible cases at equiprobability. First, if $W \sim a \mu^N$ (with $a>0$, $\mu>1$ and $N \equiv number \; of \; elements \rightarrow \infty$), then $S_1=\ln W \sim (\ln \mu)N \propto N$, hence $q^*=1$. Second, if $W \sim b N^\nu$ (with $b>0$, $\nu>0$ and $N \rightarrow \infty$), then, for $q<1$, $S_q = \frac{W^{1-q}-1}{1-q} \sim \frac{W^{1-q}}{1-q} \sim \frac{b^{1-q}}{1-q} N^{\nu(1-q)}$, consequently the choice $q^* = 1- 1/\nu$ implies $S_{q^*} \propto N$, once again recovering the traditional proportionality between entropy and $N$, when $N>>1$. 

For a system in thermal equilibrium with a thermostat (canonical ensemble), we must optimize $S_q$ with a further restriction, namely \cite{tsamepla}
\begin{equation}
\sum_{i=1}^W P_i \epsilon_i= U_q \;,
\end{equation} 
where the escort distribution $P_i  \equiv p_i^q / \sum_{j=1}^W p_j^q$, $\{\epsilon_i\}$ are the eigenvalues of the corresponding Hamiltonian (with the associated boundary conditions), and $U_q$ is a fixed finite value for the generalized internal energy. The optimizing distribution is then given by
\begin{equation}
p_i = \frac{e_q ^{-(\epsilon_i-U_q)/T_q}}       {\sum_{j=1}^W e_q ^{-(\epsilon_j-U_q)/T_q  }}\;,
\end{equation} 
where $T_q \equiv T \sum_{i=1}^W p_j^q\;$, $1/T \equiv \beta$ being the Lagrange parameter associated with constraint (3). We can verify that, for all values of $q$, $p_i$ is invariant under shifts of the zero of the spectrum of energies $\{\epsilon_i\}$. Also, for $q=1$, we recover the usual BG equilibrium distribution $p_i = e ^{-\epsilon_i/T}  /\sum_{j=1}^W e ^{-\epsilon_j/T  }$. Eq. (4) can be rewritten in another form, namely
\begin{equation}
p_i = \frac{e_q ^{-\epsilon_i/T_q^{\prime}}}       {\sum_{j=1}^W e_q ^{-\epsilon_j/T_q^{\prime}  }}\;,
\end{equation} 
with $T_q^{\prime} \equiv T_q + (1-q) U_q$. For fixed and large values of $T_q^{\prime}$, we have $p_i \sim (1-\epsilon_i/T_q^{\prime})/[\sum_{j=1}^W (1-\epsilon_j/T_q^{\prime})]$ {\it for all values of $q$}. In other words, all nonextensive equilibrium thermostatistics share, at large temperatures, a common distribution which can be equivalently considered to be the BG one \cite{cohentannoudji}.

These are the essential steps. From these, many other have been developed for arbitrary $q$, such as the $H$-theorem, Ehrenfest theorem, Bogolyubov inequality, factorization of the likelihood function, Onsager's reciprocity theorem, Kramers and Wannier relation, Pesin theorem (conjecture), fluctuation-dissipation theorem, and others. Also, a variety of traditional statistical-mechanical techniques for treating many-body systems is now available for arbitrary $q$, such as Green functions, variational method, perturbative methods, path integrals, Lie-Trotter formula, simulated annealing, and others. For review of all these, see \cite{tsallisbjp,tsallisspringer,tsallisdenton} and references therein.

Finally, in Table 1, we present the most relevant historical steps of the foundations of BG statistical mechanics, and their generalizations for arbitrary $q$. In 1860, Maxwell presented its celebrated Gaussian distribution of velocities \cite{maxwell}. In 1872, Boltzmann \cite{boltzmann} arrived, through the {\it molecular chaos hypothesis} ({\it stosszahlansatz}), to the celebrated exponential weight as the stationary state of his partial derivative kinetic
equation for distributions in the so called $\mu$-space (projection, on the one-particle phase space, of the states of all particles). In 1902, Gibbs \cite{gibbs} presented how, within a variational principle using entropy as the relevant functional, the exponential weight can be reobtained, this time in a more general framework, namely in the so called $\Gamma$-space (phase space of all particles). Gibbs equilibrium distribution was later reobtained through a variety of other manners, namely by Darwin and Fowler in 1922 \cite{darwinfowler} using a steepest descent argument, by Khinchin in 1949 \cite{khinchin1} using the law of large numbers, by Balian and Balazs in 1987
\cite{balian} and by Kubo et al in 1988 \cite{kubo}, performing countings in the microcanonical ensemble (isolated system). In parallel with these developments, Shannon in 1948 \cite{shannon} and Khinchin in 1953 \cite{khinchin2} established necessary and sufficient conditions for the entropic functional to be
$-\sum_{i=1}^W p_i \ln p_i$. All these arguments have been generalized for arbitrary $q$. Gibbs path was followed in 1988 and thereafter \cite{tsallisjsp,curado,tsamepla}, the Darwin-Fowler, Khinchin and Balian-Balazs paths were followed in 2000 by Abe and Rajagopal in \cite{ardarwinfowler}, \cite{arkhinchin} and \cite{arbalian} respectively, and the Kubo path was followed by Abe and Rajagopal in 2001 \cite{arkubo}. Boltzmann path was followed in 2001 by Lima, Silva and Plastino \cite{lima} as well as by Kaniadakis in  \cite{kaniadakis}.
Shannon and Khinchin paths for the necessary and sufficient for the entropic form were respectively followed by Santos in 1997 \cite{santos} and
by Abe in 2000 \cite{abekhinchin}. All these generalizations, without exception, proved to be consistent among them and consistent with Eqs. (1--4). 

At this point it seems appropriate to comment that, as a whole, the formalism exhibits a remarkable mathematical ``texture", which quite naturally extends to arbitrary values of $q$ the properties that are since long known for $q=1$. 

\noindent
{\it {\bf Theoretical physics}}

Let us now argue in the sense of transforming the above mathematical formalism into theoretical physics by comparing theoretical with experimental results. We must however warn the reader that many of the available applications of nonextensive statistical mechanics concern open systems ({\it stationary states} of open systems), and not only {\it (meta)equilibrium states} of the time-independent Hamiltonian systems addressed in Eqs. (3--5). To be more precise, all thermodynamical equilibrium states are stationary solutions of some family of partial derivative equations (e.g., of the Boltzmann kinetic equation, for classical systems), but the opposite is not true. There are stationary
states which are not tractable in thermostatistical terms, this is to say susceptible of being founded in geometrical aspects of some phase space (or of some Riemann subspace in that phase space) or analogous spaces such as the Hilbert or Fock ones. To illustrate what we mean about geometrical founding, it is certainly instructive to recall that it is the symmetrization or the antisymmetrization of the corresponding wave functions that determines the
transmutation of Maxwell-Boltzmann statistics into Bose-Einstein and Fermi-Dirac ones, consistently leading to entropic functionals which differ from the classical one. The point is that if the dynamics of the system is completely known, its possible stationary states always are in principle calculable, but for some classes of such systems {\it it is not necessary to follow its dynamics}: we can directly, geometrically, calculate the stationary state, which can then be considered as optimizing some entropic functional. Statistical mechanics focuses on such states. The difficulty is of course to know {\it a priori} what specific statistics is to be applied to a given system. We come
back onto this point later on. But at the present stage, let us emphasize that in our understanding {\it statistical mechanics emerges if and only if dynamics can in some sense be replaced by geometry}. Unfortunately it is by no means clear {\it when} this is possible, but we shall refer to it as {\it geometrizable dynamics}. As an attempt to clarify these arguments, we schematically display in Figs.~1 and 2 respectively the traditional and the present views on the connections between dynamics and statistics (see also Cohen's contribution to the present proceedings). In Fig. 1 it is stressed that, in the traditional view, an unique type of thermal equilibrium exists and this is the BG one (either in the micro-canonical, the canonical or the various grand-canonical forms). We believe that statistical mechanics is wider than
that, as indicated in Fig. 2. Every time that dynamics (of a finite or infinite system) can be automatically taken into account by geometrical considerations, theory of probabilities can, for a variety of purposes, efficiently replace the knowledge of the time evolution of the mechanical system. Within classical BG statistical mechanics, dynamics of vast classes of isolated systems (typically involving short-range interactions) can be replaced by the hypothesis of equiprobability in the occupation of the accessible phase space together with a connection of macroscopic entropy with the relevant phase space volume. We are of course referring to the Boltzmann principle $S=k \ln W$ (see also Gross' contribution to the present proceedings). It is however physically appealing to think that more subtle situations can also be handled within
statistical-mechanical methods. Such could be the case when, in addition to the knowledge of the volume $W$, we need to characterize a physical, dynamical {\it bias} : this is the role of $q$ within nonextensive statistical mechanics. The just mentioned Boltzmann principle would have to be generalized in such cases by $S_q = k [W^{1-q} -1]/[1-q]$. Of course, there is no reason for thinking that such {\it geometrization} of dynamics could not be in principle done for other, possibly more complex, systems, {\it outside} of the $q$-statistics focused
here. Within this scenario, nonextensive statistical mechanics appears to be just the first non-Boltzmannian thermostatistical formalism; many others are in principle thinkable, corresponding to various manners for replacing (specific classes of anomalous) dynamics by geometry. In our present formalism it is clear that $W$ is a geometrical concept, but the reader might be puzzled by the fact that we are including $q$ in the same category. This point will become transparent soon, when we shall illustrate how multifractal geometrical considerations enable us, at least for some simple specific cases, to uniquely determine $q$ {\it a priori} from the mechanical characterization of the system (and {\it not only} from fitting procedures, as frequently done, {\it faute de mieux}, in the experimental literature).  

To make connection with nature, it is mandatory to mention that the present formalism has been successfully applied to a considerable variety of systems, such as L\'evy \cite{levy} and correlated \cite{correlated} anomalous diffusions, peculiar velocities in spiral galaxies \cite{galaxy}, turbulence in electron plasma \cite{bogho}, fully developed turbulence \cite{ramos,turbulence,turbulencebeck,swinney}, citations of scientific papers
\cite{citations}, linguistics \cite{montemurro}, reassociation in folded proteins \cite{bemski}, quantum entanglement \cite{rajagopal,sethmichel}, electron-positron annihilation \cite{bediaga}, quark-gluon plasma \cite{rafelski}, cosmology \cite{ivano,oliveirasoares}, hadronic scattering \cite{ion}, motion of {\it Hydra viridissima} \cite{arpita}, low-dimensional maps \cite{maps,moura}, inertial classical planar rotators ferromagnetically coupled at long distances \cite{cataniario,cataniario2}, among others.   
To be more precise, this formalism addresses systems which, in one way or another, exhibit some relevant multifractal structure. This can occur through a variety of physical mechanisms, such as spatial and/or temporal long-range interactions, mesoscopic dissipation, multifractal boundary conditions, quantum entanglement, and others. 

This is a good point for warning the reader about the fact that {\it stretched exponentials} and {\it $q$-exponentials} can be numerically very close to each other for intermediate values of the abscissa (see Fig. 3), although they are definitively very different both close to the origin and approaching infinity. It is important to realize these features in order to really appreciate the strength of experimental and computational evidences favoring one or the other
functional form whenever fittings are involved. The present formalism naturally leads to $q$-exponentials, rather than to stretched exponentials (frequently used for fittings during the last 10--15 years). But only fittings that are satisfactory over relatively large physical ranges can be acceptable in order to distinguish between these two functional forms. For example, in the case of {\it Hydra viridissima} just mentioned \cite{arpita}, both a $q$-exponential and a stretched exponential fit well the experimental distribution of velocities if only relatively small velocities are taken into account. In this particular case, it is because large velocities were experimentally measured as well, that it became possible to dismiss the stretched exponential function, and retain the $q$-exponential one.

\noindent
{\it {\bf Closed theory}}

We shall from now on note $q_{eq}$ the value of $q$ characterizing the equilibrium distribution (or the stationary distribution if the system is open); typically $q_{eq} \ge1$ (see, for instance, \cite{turbulencebeck,swinney,bediaga,rafelski,ivano,oliveirasoares,arpita,cataniario2}). Let us also address a different quantity, noted $q_{mix}$, related to the mixing properties of the system; typically $q_{mix} \le 1$ (see, for instance, \cite{turbulence,maps}). The relationship between these two different values of $q$ is under intensive study nowadays (see \cite{ernestogarin}). Our focus here is, as mentioned previously, to develop a scenario where $q_{mix}$, $q_{eq}$ and the metaequilibrium states that frequently emerge for some nonextensive systems, play deeply entangled roles. The final outcome is to illustrate the lines along which dynamics can a priori determine the values of $q$ to be used for specific physical models, so that the whole thermostatistical formalism becomes a closed and complete theory.

To start uncovering the scenario it is enough to consider systems whose phase space (noted $x$) is one-dimensional, e.g., one-dimensional maps such as the logistic one. If we note $\Delta x(t)$ the discrepancy, as a function of time, of two trajectories initially discrepant of $\Delta x(0)$, we can define the {\it sensitivity to the initial conditions} (or {\it mixing} function) $\xi(t) \equiv \lim_{\Delta x(0) \rightarrow 0} \Delta x(t)/\Delta x(0)$. The most frequent case is that where $\xi$ satisfies the following differential equation:
\begin{equation}
\frac{d\xi}{dt} = \lambda_1 \;\xi\;,
\end{equation}  
where $\lambda_1$ is the Lyapunov exponent (the subscript 1 will become clear soon). It follows that
\begin{equation}
\xi = e^{\lambda_1 \; t}
\end{equation}  
Positive and negative values for $\lambda_1$ respectively correspond to {\it strong} sensitivity and insensitivity to the initial conditions. What happens when $\lambda_1=0$? The generic answer is that Eq. (6) is not applicable anymore, and we must take into account the next term, so we consider now
\begin{equation}
\frac{d\xi}{dt} = \lambda_{q_{mix}} \;\xi^{q_{mix}}\;,
\end{equation}  
where we focus on the case $ \lambda_{q_{mix}}>0$ and $q_{mix} \le 1$. The solution now is 
\begin{equation}
\xi = [1+(1-q_{mix})\; \lambda_{q_{mix}}\;t]^{1/(1-q_{mix})} \;,
\end{equation}  
which reproduces solution (7) if $q_{mix}=1$, and asymptotically behaves like $t^{1/(1-q_{mix})}$ if $q_{mix} <1$. The latter will be referred to as {\it weak} sensitivity to the initial conditions (or {\it weak} mixing). Let us finally consider the most general case along this line, namely
\begin{equation}
\frac{d\xi}{dt} =\lambda_1 \; \xi + (\lambda_{q_{mix}}-\lambda_1) \;\xi^{q_{mix}}\;,
\end{equation}  
whose solution is
\begin{equation}
\xi = \Bigl[1- \frac{\lambda_{q_{mix}}}{\lambda_1} + \frac{\lambda_{q_{mix}}}{\lambda_1}e^{(1-q_{mix})\lambda_1 t} \Bigr]^{\frac{1}{1-q_{mix}}} \;.
\end{equation}  
This function is illustrated in Fig.~4 in such a way as to exhibit the crossover from the power-law regime at intermediate times to the exponential regime at long times, which occurs when $0<\lambda_1<<\lambda_{q_{mix}}$ and $q_{mix}<1$. More precisely, for times $t$ satisfying $0 \le t <<t^* \equiv 1/[(1-q_{mix})\; \lambda_{q_{mix}}]$, we have an integrable-like regime (with $\xi \sim 1 + \lambda_{q_{mix}}\;t$) (i.e, characterized by $q=0$), for times satisfying $t^* <<t <<t^{**} \equiv   1/[(1-q_{mix})\; \lambda_1]$ we have a power-law regime (with $\xi \sim [(1-q_{mix})\; \lambda_{q_{mix}}\;t]^{1/(1-q_{mix})}$, i.e., characterized by $q=q_{mix}$), and finally for times satisfying $t>>t^{**}$ we have an exponential regime (with $\xi \sim \frac{\lambda_{q_{mix}}}{\lambda_1}\;e^{\lambda_1 \; t}$, i.e., characterized by $q=1$). These facts lead to the following nonuniform convergence:  $\lim_{\lambda_1 \rightarrow 0} \lim_{t \rightarrow \infty} \frac{\ln \xi}{\ln t} = \lim_{\lambda_1 \rightarrow 0} \lim_{t \rightarrow \infty} \frac{\lambda_1\;t}{\ln t}=\infty$, whereas $\lim_{t \rightarrow \infty} \lim_{\lambda_1 \rightarrow 0}  \frac{\ln \xi}{\ln t} = \lim_{t \rightarrow \infty} \frac{1}{1-q_{mix}}\frac{\ln t}{\ln t} = \frac{1}{1-q_{mix}} < \infty$. Analogously, using the $q$-logarithm function $\ln_q x$, we have $\lim_{\lambda_1 \rightarrow 0} \lim_{t \rightarrow \infty} \frac{\ln_{q_{mix}} \xi}{t} =\lim_{\lambda_1 \rightarrow 0} \lim_{t \rightarrow \infty} \frac{e^{\lambda_1 t}/(1-q_{mix})}{t} = \infty$, whereas $\lim_{t \rightarrow \infty} \lim_{\lambda_1 \rightarrow 0}  \frac{\ln_{q_{mix}} \xi}{ t} =\lambda_{q_{mix}} < \infty$. In other words, for $1<<t << t^{**}$, we have $\frac{\ln_{q_{mix}} \xi}{t} \simeq \lambda_{q_{mix}}$, whereas, for $t >> t^{**}$, we have $\frac{\ln \xi}{t} \simeq \lambda_1$. We believe that these features constitute the basic scenario of validity of Boltzmann-Gibbs statistical mechanics versus validity of nonextensive statistical mechanics. This belief is supported by at least three examples, namely the standard map (see \cite{baldovin} and references therein), another, billiard-inspired, two-dimensional conservative map \cite{fulviocasati}, and the  system of $N$ classical inertial planar rotators coupled all with all through long-range interactions (so called $\alpha$-XY model, which for $\alpha=0$ reproduces the HMF model; see \cite{cataniario,cataniario2} and references therein).

The standard map is a conservative map with a two-dimensional phase space (the lowest dimension at which a map can be conservative). It includes a nonlinear term introduced by a constant $a$. In the limit $a=0$ the system is integrable, and for $a \ne 0$ it is chaotic (i.e., it has over entire regions of the phase space two nonzero Lyapunov exponents of opposite sign and equal in absolute value). The positive Lyapunov exponent $\lambda_1$ is a monotonic function of $a$ which vanishes for $a=0$. In the limit $0 < a <<1$, this map precisely exhibits \cite{baldovin} the scenario described above with $q_{mix} \simeq 0.3$ as studied through the time evolution of the entropic form $S_q$. Indeed, for times below a crossover time (which appears to diverge when $a$ approaches zero), $S_{0.3}$ increases linearly with time (see \cite{baldovin} for details), whereas for times above that crossover time, it is $S_1$ which linearly increases with time. The same behavior is observed for the above mentioned billiard-like map, but with $q_{mix} \simeq 0.5$.

Let us now turn onto our third example. The system of rotators mentioned above has been studied numerically and presents, in the microcanonical ensemble (i.e., isolated) a second order phase transition at some total energy (conveniently scaled with $N$). This system presents anomalies both above and below that critical energy. Above that point, the entire Lyapunov spectrum collapses to zero when $N \rightarrow \infty$. This is to say, $1/N$ plays a role analogous
to $a$ in the map just described. Below that critical point, the system exhibits at least two (probably only two) basins of attraction with respect to the initial conditions at which the system is dynamically started. There is a basin of attraction which exhibits an equilibrium at the temperature that the recipe of BG statistical mechanics provides, and whose distribution of velocities precisely is the Maxwellian one. But if we start from the other basin of attraction, the system equilibrates at a finite temperature {\it different} (below) from that indicated by BG statistics, and its distribution of velocities exhibit unusual long tails. After some long time the system crosses over
essentially to the BG equilibrium state. The duration of this anomalous metastable state diverges with $N$, in a manner which, once again, strongly reminds the scenario we advanced in the present paper. Indeed, if we consider the $\lim_{N \rightarrow \infty} \lim_{t \rightarrow \infty} $ case, the equilibrium appears to be correctly described within BG statistical mechanics,
but if we consider the $\lim_{t \rightarrow \infty} \lim_{N \rightarrow \infty}  $ case, this is definitively not true, and some other thermostatistical description becomes necessary, apparently the nonextensive one. To make the scenario stronger, it can be verified that, during this metastable state, the Lyapunov spectrum also vanishes in the limit $N \rightarrow \infty$.

We have sketched above what happens with mixing. What can we say about equilibration and $q_{eq}$? In other words, how physical quantities relax onto the corresponding equilibrium values? Let us refer once again to the logistic map as an illustration. In the region where strong chaos exists (hence above the chaos threshold), for fixed $W$ and using an ensemble where all initial conditions belong to a {\it single} among the $W$ windows of the partition, $S_1(t)$ exhibits a linear increase with time and then a saturation at $S_1(\infty)$. Moreover we verify that 
$\sigma_1 \equiv |1 -   \frac{S_1(t)}{S_1(\infty)}    | \sim e^{-t/\tau_1}$,
$\tau_1$ being of the order of $1/\lambda_1$ (consistently with Krylov's ideas more than half a century ago \cite{krylov}). In other words, $\sigma_1$ essentially satisfies $\frac{d\sigma_1}{dt} = -\sigma_1/\tau_1$. However, at the chaos threshold, $\tau_1$ diverges, and the time evolution of $\sigma_{q_{mix}} \equiv |1 - \frac{S_{q_{mix}}(t)}{S_{q_{mix}}(\infty)}|$ is essentially given \cite{ernestogarin} by $\frac{d\sigma_{q_{mix}}}{dt} = -(\sigma_{q_{mix}})^{q_{eq}}/\tau_{q_{eq}}$, with $q_{eq}>1$ and $\tau_{q_{eq}}$ hopefully of the order of $1/\lambda_{q_{mix}}$. The  solution of this differential equation is given by $\sigma_{q_{mix}} = 1/[1+(q_{eq}-1)\;t/\tau_{q_{eq}}]^{[1/(q_{eq}-1)]}$, which reproduces $e^{-t/\tau_1}$ for $q_{eq}=1$. Furthermore, $q_{eq}$ depends on $W$, and, in the $W \rightarrow \infty$ limit ({\it infinitely fine graining}), we observe (within some degree of accuracy) \cite{ernestogarin} the following finite size scaling
\begin{equation}
q_{eq}(\infty)-q_{eq}(W) \propto \frac{1}{W^{q_{mix}}}\;,
\end{equation}
with $q_{eq}(\infty)>1$. This fascinating relation has up to now been verified only for the $z$-logistic maps. For these maps and all values of $z$ that have been checked,  the values for $q_{eq}(\infty)$ precisely coincide -- supreme suggestion of correctness of the present conjectural scenario! --, with the values numerically obtained in \cite{moura}, where the initial conditions were {\it spread uniformly all over the entire accessible phase space} (in a typically Gibbsian manner).  

We are unfortunately not in position to rigorously prove the validity of the above relations nor discuss the detailed hypothesis they must involve. It has however been possible to give some physical consistency to the conjectural scenario by analyzing three different phenomena, namely related to electron-positron annihilation \cite{bediaga}, to fully developed turbulence \cite{swinney} and to the Henon-Heiles Hamiltonian \cite{oliveirasoares}. 

The distributions of transverse momenta of hadronic jets produced by electron-positron annihilation experiments at CERN have been discussed by Bediaga et al in \cite{bediaga}. The values for $q_{eq}$ they obtained depend on the collision center-of-mass energy $E$, which plays a role similar to $W$ since the graining is finer for increasingly high energies. The data in their Table 1 can be organized as indicated in the present Fig.~5, with $q_{eq}(\infty) \simeq 1.30$, and the role played by the exponent 1/2 is that of $q_{mix}$ in Eq. (12).

The distributions of velocity differences in fully developed turbulence in Couette-Taylor experiments have been discussed by Beck et al in \cite{swinney} for four typical values of the Reynolds number $Re$ and millions of experimental distances $r$ (in units of the Kolmogorov length $\eta$). The data they present in their Fig.~3 can be organized as shown in the present Fig.~6, with $q_{eq}(\infty) \simeq 1.45$, and the role played by the exponent 0.37 is that of $q_{mix}$ in Eq. (12). Incidentally, 0.37 is the value used by the Arimitsu's \cite{turbulence} to produce (in one of their calculations) such distributions of velocity differences. It is worthy mentioning also that the value 1.45 is very close to $3/2$ recently advanced by Beck \cite{becklagrangian} for Lagrangian turbulence.

Saddle-point dynamics of the Henon-Heiles system have been discussed by Soares et al \cite{oliveirasoares}. In their Table 1 they show the numerical values obtained for $\gamma \equiv 1/(q_{eq}-1)$ as a function of the control parameter $k$, known to play a role analogous to a Reynolds number, hence $k$ characterizes $W$ and the $k \rightarrow \infty $ limit corresponds to infinitely fine graining. The data presented in that Table can be organized as shown in the present Fig.~7, thus giving support to relation 
(12), with $q_{eq}(\infty) \simeq 2.81$, and the role played by the exponent 0.35 is that of $q_{mix}$.

Summarizing, the basic picture which emerges from all the above considerations is as follows. If the system is strongly chaotic (in the sense that its spectrum of Lyapunov exponents includes positive values, i.e., exponential mixing), then the measure of ignorance (entropy) to be used is, as well known, $S_1$, from which the entire Boltzmann-Gibbs statistical mechanics can be derived. If, however, the system is only weakly chaotic (zero Lyapunov exponent spectrum, and power-law mixing), then several indications exist which suggest that we should instead use $S_{q_{mix}}$ in what concerns a {\it finite} entropy production,
and  $S_{q_{eq}}$ to discuss the corresponding equilibrium thermodynamics, both $q_{mix}$ and $q_{eq}$ being uniquely determined by the dynamics of the specific physical system (to be more precise, the graining degree, characterized here by $W$, also enters in the determination of $q_{eq}(W)$; it is the value $q_{eq}(\infty)$ the one which only depends on the fundamental dynamics). As we
see, in this scene there are two protagonists, namely mixing and equilibration. Although deeply entangled, they are different concepts, typically represented by two different values of $q$, one of them ($q_{mix}$) not above unity, and the other one ($q_{eq}$) not below unity. These two values of $q$ merge on a single one ($q=1$) for Boltzmann-Gibbs thermostatistics (thus transforming power laws in exponentials). It is certainly allowed to think that this is perhaps at the origin of not few of the {\it warm} controversies in our community about the foundations of statistical mechanics!

One of us (CT) acknowledges very fruitful discussions with E.G.D. Cohen,  E.M.F. Curado, A.C.N. de Magalhaes, A. Plastino, A.R. Plastino (see Appendix), D. Prato, I.D. Soares, H.L. Swinney, V.V. Uchaikin (see Appendix) and R.O. Vallejos. We are grateful to C. Beck et al for kindly providing to us the numerical data corresponding to Fig.~3 of \cite{swinney}.  Finally, unforgettable hospitality from the main organizers of the meeting, namely P. Quarati, M. Lissia and A. Rapisarda, is also warmly acknowledged. This work has benefited from partial support from CNPq, PRONEX, CAPES and FAPERJ (Brazilian agencies).\\

\noindent
{\it {\bf APPENDIX}}

This Appendix focuses on an interesting point that was raised during the meeting by V.V. Uchaikin, and reports A.R. Plastino' s related remarks. 

Let us first remind, along the lines of \cite{lubbe} (see also \cite{cover}), the $q=1$ case. Let us assume that we have a continuous probability distribution $p(x)$ (with $ \int dx \; p(x)=1$, $x$ being a one-dimensional real variable) and, using the entropic functional $S_1[p(x)] = -\int dx \; p(x)\; \ln p(x)$, we wish to consider its discretization. In other words, we consider $p_i = p(x_i) \Delta$, where $\Delta$ represents a graining for $x$, and $i=1,2,...,W$. It follows that, in the limit $\Delta \rightarrow 0$, $S_1[p(x)] \sim - \sum_{i=1}^W p(x_i) \ln \frac{p(x_i)}{\Delta} =  - \sum_{i=1}^W p(x_i) \ln p(x_i) + \ln \Delta$, hence,
\begin{equation}
S_1[p(x)] \sim S_1(\{p(x_i)\}) + \ln \Delta\;.
\end{equation}
We see that $\Delta \rightarrow 0$ leads to a divergence. This divergence is not surprising and corresponds \cite{lubbe} to the fact that it is necessary an {\it infinite}  number of yes/no answers to resolve the uncertainty associated with a continuous distribution. In practice, it is chosen $\Delta=1$, therefore the continuous and discrete versions of the entropy provide the same result. The choice $\Delta =1$ corresponds, when we consider the passage from quantum to classical statistical mechanics, to the measure of phase space in units of $\hbar$ for every pair of conjugate mechanical variables.

Let us now address this point for arbitrary $q$. Using the entropic functional $S_q[p(x)] \equiv \frac{1-\int dx \; [p(x)]^q}{q-1}$ we straightforwardly obtain
\begin{equation}
S_q[p(x)] \sim \Delta^{1-q}S_q(\{p(x_i)\}) + \ln_q \Delta\;,
\end{equation}
where we have used Eq. (1). Of course, this equation recovers Eq. (13) for $q=1$. We see that, on top of an extra {\it additive} term, like in the $q=1$ case, we have here an extra {\it multiplicative} term. Nevertheless, like in the $q=1$ case, the choice $\Delta=1$ makes the continuous and discrete versions of the nonextensive entropy to coincide.

\begin{table}
\begin{center}
ENTROPIC FORM AND EQUILIBRIUM STATISTICS: FOUNDATIONS
\end{center}
\def\baselinestretch{1}
\begin{tabular}{|c|c|c|}
& BG
& \multicolumn{1}{c|}{$q\neq 1$} \\  \hline
 distribution of velocities
&  \multicolumn{1}{l|}{Maxwell 1860}
& \multicolumn{1}{l|}{R.S. Mendes and C. Tsallis} \\
 at equilibrium
&
&\multicolumn{1}{r|}{Phys Lett A {\bf 285}, 273 (2001)}  \\  \hline
molecular chaos hypothesis
&  \multicolumn{1}{l|}{Boltzmann 1872}
& \multicolumn{1}{l|}{J.A.S. Lima,R. Silva and A.R. Plastino} \\
({\it Stosszahlansatz})
&
&\multicolumn{1}{r|}{Phys Rev Lett {\bf 86}, 2938 (2001)} \\
&
&\multicolumn{1}{l|}{G. Kaniadakis } \\
&
&\multicolumn{1}{r|}{Physica A {\bf 296}, 405 (2001)} \\ \hline
optimization of entropy
&  \multicolumn{1}{l|}{Gibbs 1902}
& \multicolumn{1}{l|}{C.Tsallis}\\
with constraints
&
&\multicolumn{1}{r|}{ J Stat Phys {\bf 52}, 479 (1988)} \\
&
&\multicolumn{1}{l|}{E.M.F. Curado and C.Tsallis} \\
&
&\multicolumn{1}{r|}{J Phys A {\bf 24}, L69 (1991)} \\
&
&\multicolumn{1}{l|}{C.Tsallis, R.S. Mendes and A.R. Plastino} \\
&
&\multicolumn{1}{r|}{Physica A {\bf 261}, 534 (1998)} \\ \hline
steepest descent
&  \multicolumn{1}{l|}{Darwin-Fowler 1922}
& \multicolumn{1}{l|}{S. Abe and A.K. Rajagopal} \\
&
&\multicolumn{1}{r|}{J Phys A {\bf 33}, 8733 (2000)} \\  \hline
conditions of uniqueness
&  \multicolumn{1}{l|}{Shannon 1948}
&\multicolumn{1}{l|}{R.J.V. Santos} \\
of the entropy
&
&\multicolumn{1}{r|}{J Math Phys {\bf 38}, 4104 (1997)} \\  \hline
law of large numbers
&  \multicolumn{1}{l|}{Khinchin 1949}
& \multicolumn{1}{l|}{S. Abe and A.K. Rajagopal} \\
&
&\multicolumn{1}{r|}{Europhys Lett {\bf 52}, 610 (2000)} \\  \hline
compact conditions of
&  \multicolumn{1}{l|}{Khinchin 1953}
& \multicolumn{1}{l|}{S. Abe} \\
uniqueness of the entropy
&
&\multicolumn{1}{r|}{Phys Lett A {\bf 271}, 74 (2000)}  \\  \hline
counting in the
&  \multicolumn{1}{l|}{Balian-Balazs 1987}
& \multicolumn{1}{l|}{S. Abe and A.K. Rajagopal} \\
microcanonical ensemble
& \multicolumn{1}{l|}{Kubo et al 1988}
&\multicolumn{1}{r|}{Phys Lett A {\bf 272}, 341 (2000)} \\
&
&\multicolumn{1}{r|}{Europhys Lett {\bf 55}, 6 (2001)} \\
\end{tabular}
$ $\\
\caption{
Historical steps of the foundations of BG statistical mechanics (both equilibrium distribution and entropic form), and their $q \ne 1$ counterparts. Gibbs 1902 and Kubo et al 1988 refer not to the dates when the original works were essentially done, but rather to the books where they are reproduced.}
\end{table}

\renewcommand{\baselinestretch}{1.0}
\begin{figure}[htb]
\setlength{\epsfxsize}{11.5cm}
\centerline{\mbox{\epsffile {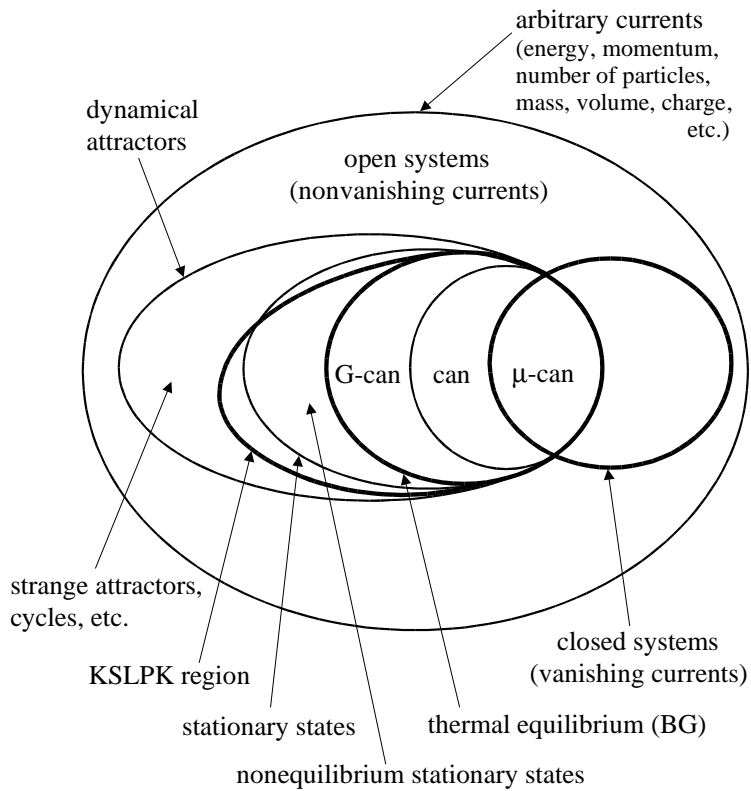}}}
\caption{\small Traditional view (schematical) of the place of statistical mechanics for classical, quantum or relativistic dynamical systems with a finite or infinite number $N$ of particles. Thermal equilibrium only occurs for conservative systems and is necessarily of the BG class, independently of the ordering of limits such as the $t \rightarrow \infty$ and the $N \rightarrow \infty$ ones. This is certainly the case of short-range interactions which present no delicate singularity at say the origin. The micro-canonical ensemble can be seen as the particular case of the canonical one when the temperature diverges; the canonical one can in turn be seen as the particular case of the grand-canonical ensemble when all chemical potentials vanish. KSLPK stands for Kolmogorov-Sinai-Lyapunov-Pesin-Krylov, thus meaning the region where, for the particular case of classical dynamical systems, the Kolmogorov-Sinai entropy is positive, the Lyapunov spectrum includes a positive branch, the Pesin identity is nontrivial, i.e. connecting nonvanishing quantities, and Krylov's emphasis on exponential mixing being essential in the foundations of BG statistical mechanics. The Sinai-Ruelle-Bowen (SRB) distributions are stationary states which can be both at or out from thermal equilibrium; the BG equilibrium distributions are particular cases.}
\end{figure}
\newpage

\begin{figure}[htb]
\setlength{\epsfxsize}{11.5cm}
\centerline{\mbox{\epsffile {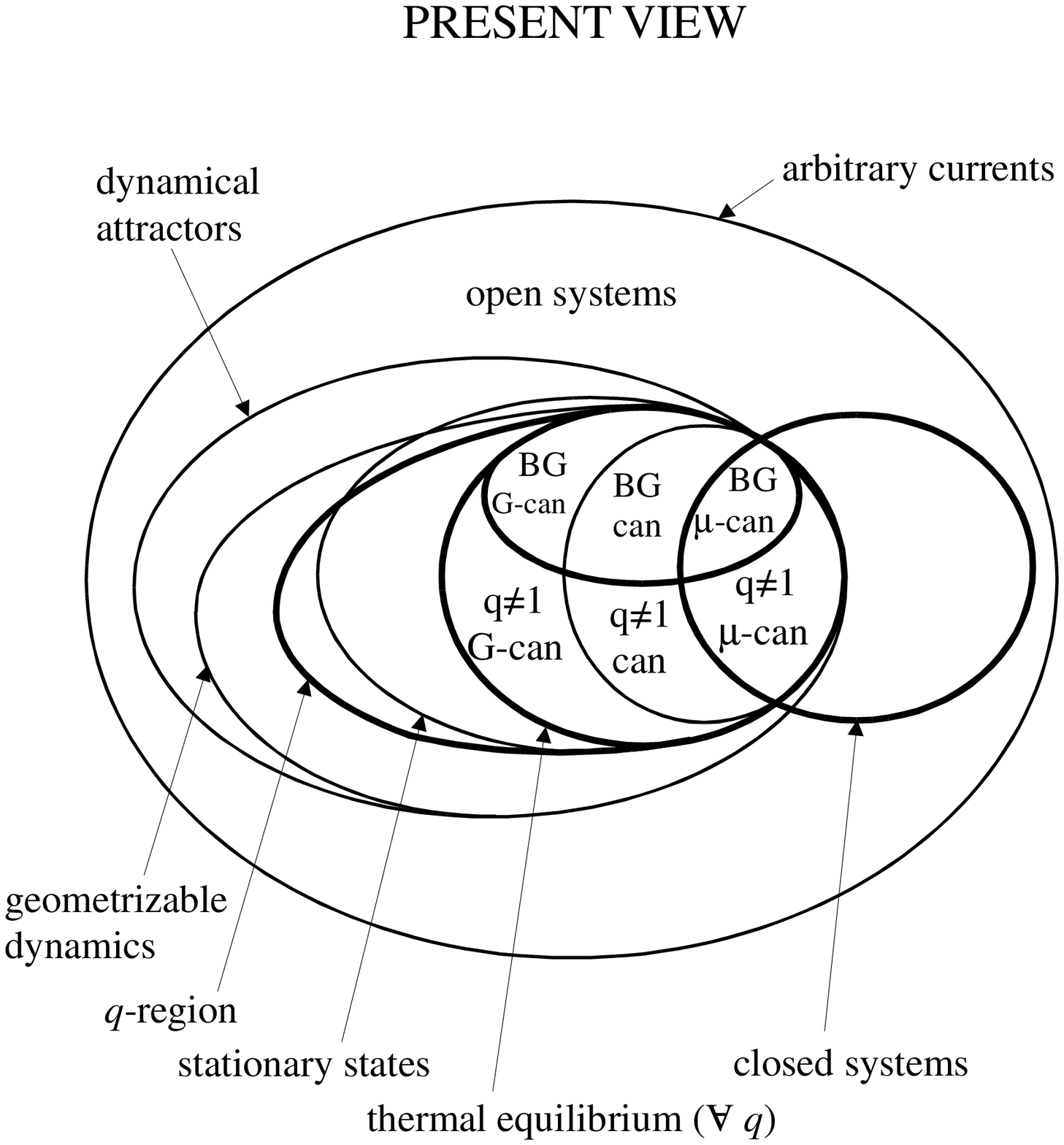}}}
\caption{\small Present view (schematical) of the place of statistical mechanics for classical, quantum or relativistic dynamical systems with a finite or infinite number $N$ of particles. The traditional ensembles are enlarged in the sense that $q$ can differ from unity. Distributions similar to those occurring at thermal $q$-equilibrium, or metaequilibrium (in the sense of metastability), can occur even for dissipative systems. Ordering of limits such as the $t \rightarrow \infty$ and the $N \rightarrow \infty$ ones can be very relevant. For example, for conservative many-body systems including long-range interactions, the  $\lim_{t \rightarrow \infty} \lim_{N \rightarrow \infty}$ ordering, which is the physical one, is non-Boltzmannian, whereas the $\lim_{N \rightarrow \infty} \lim_{t \rightarrow \infty}$, physically unobservable, corresponds to BG statistics. The $q$-region includes the 1-region, which precisely is the KSLPK region of Fig. 1.}
\end{figure}

\newpage
\begin{figure}[htb]
\setlength{\epsfxsize}{13.cm}
\centerline{\mbox{\epsffile {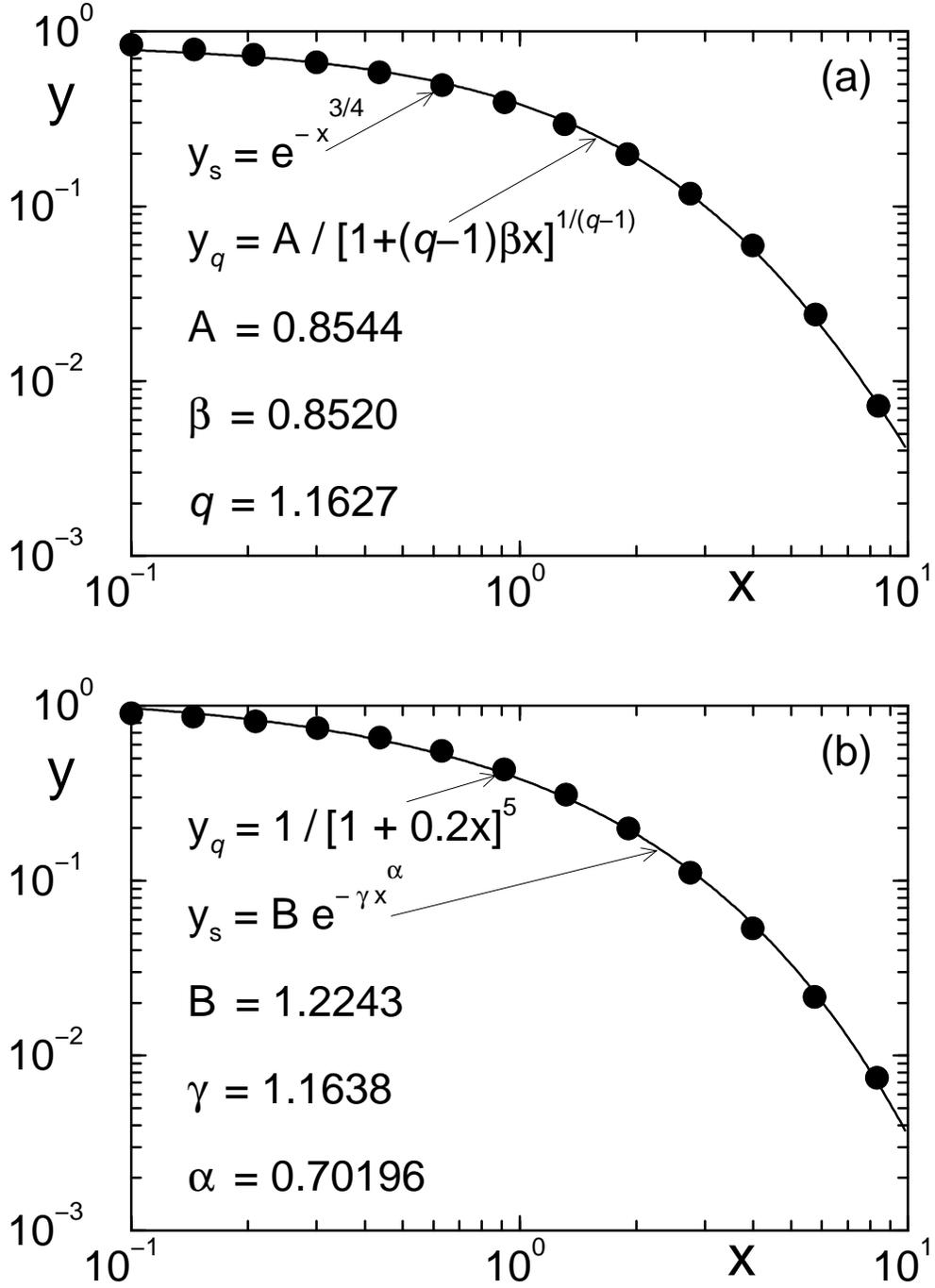}}}
\caption{\small Comparison between $q$- and stretched exponentials. (a) The circles have been calculated with a stretched exponential, and have been fitted with a $q$-exponential. (b) The circles have been calculated with a $q$-exponential, and have been fitted with a stretched exponential. The numerical discrepancies emerge only for $x<<1$ and for $x>>1$.}
\end{figure}

\newpage
\begin{figure}[htb]
\setlength{\epsfxsize}{13.cm}
\centerline{\mbox{\epsffile {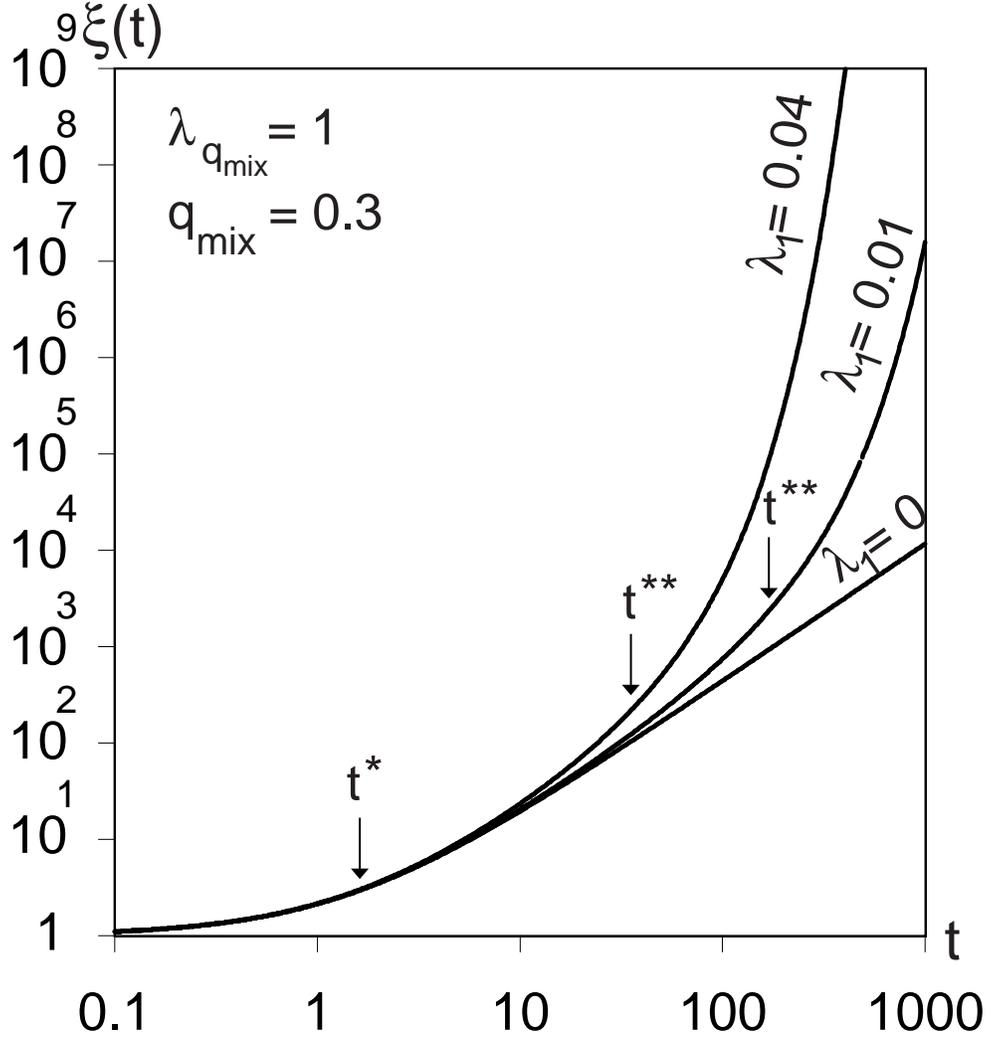}}}
\caption{\small Time evolution of the sensitivity to the initial conditions $\xi$. The crossover occurring in the limit $\lambda_1 \rightarrow 0$ becomes apparent: the smaller $\lambda_1$ is compared with $\lambda_{q_{mix}}$, the larger is the domain of validity of the power law $\xi \propto t ^{1/(1-q_{mix})}$. The crossover time $t^\star$ does not depend on $\lambda_1$; the crossover time $t^{**}$ depends on $\lambda_1$ and diverges when $\lambda_1$ vanishes.}
\end{figure}

\newpage
\begin{figure}[htb]
\setlength{\epsfxsize}{13.cm}
\centerline{\mbox{\epsffile {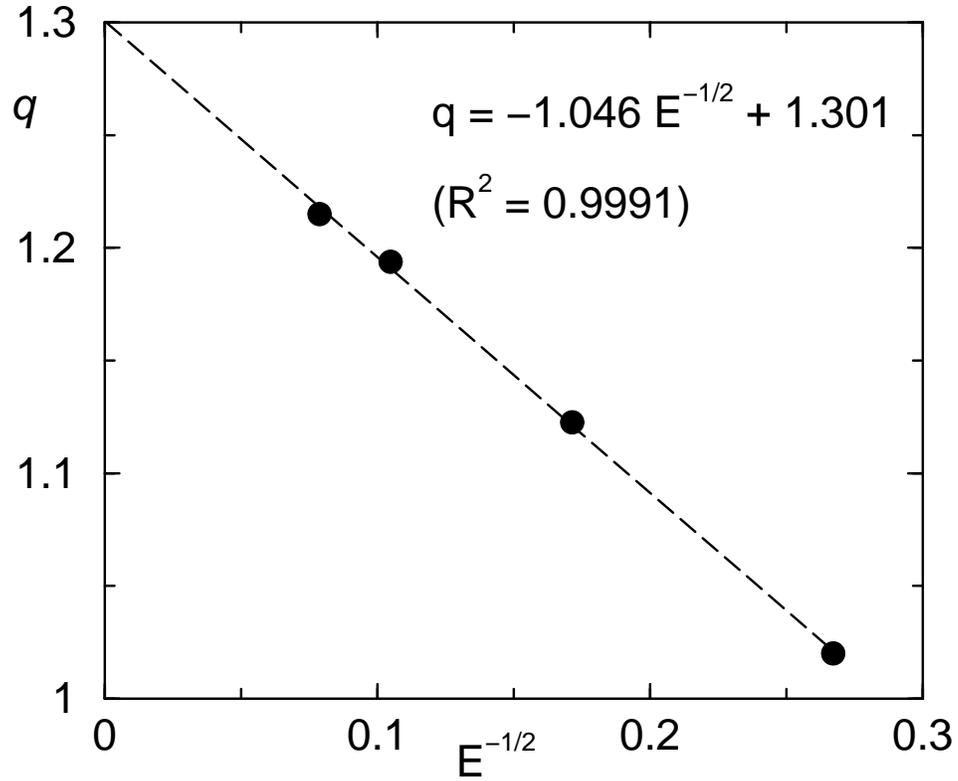}}}
\caption{\small Values of $q_{eq}$ obtained by Bediaga et al through the analysis of distributions of hadronic transverse momenta in electron-positron experiments.  We have chosen the abscissa in such a way as to produce a linear form. $R^2$ is the square linear correlation factor.}
\end{figure}

\newpage
\begin{figure}[htb]
\setlength{\epsfxsize}{13.cm}
\centerline{\mbox{\epsffile {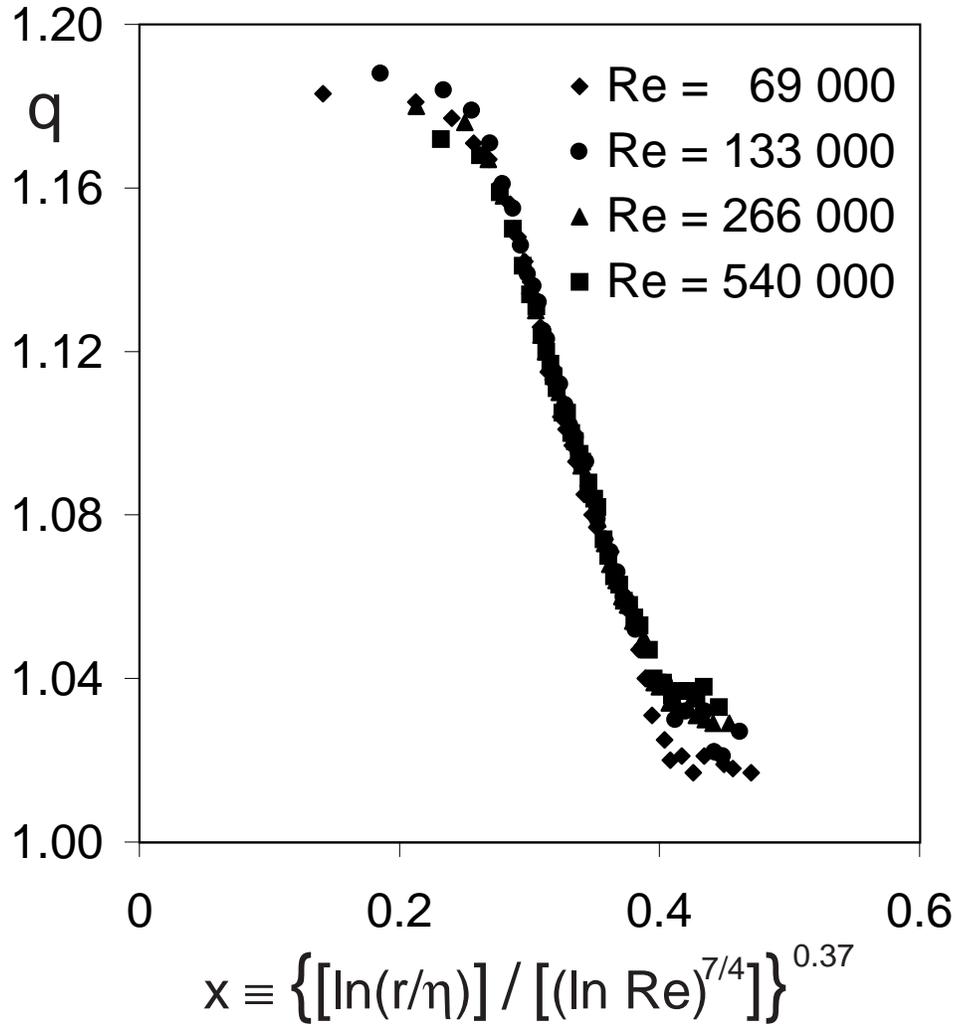}}}
\caption{\small Values of $q_{eq}$ obtained by Beck et al through the analysis of distributions of velocity differences in Couette-Taylor experiments. We have scaled $\ln (r/\eta)$ with $(\ln Re)^{7/4}$ in order to produce data collapse for different values of $Re$. Also, we have chosen the exponent 0.37 in such a way as to produce a linear form for intermediate distances. The error bar on this exponent is of the order of 0.1 .}
\end{figure}

\newpage
\begin{figure}[htb]
\setlength{\epsfxsize}{13.cm}
\centerline{\mbox{\epsffile {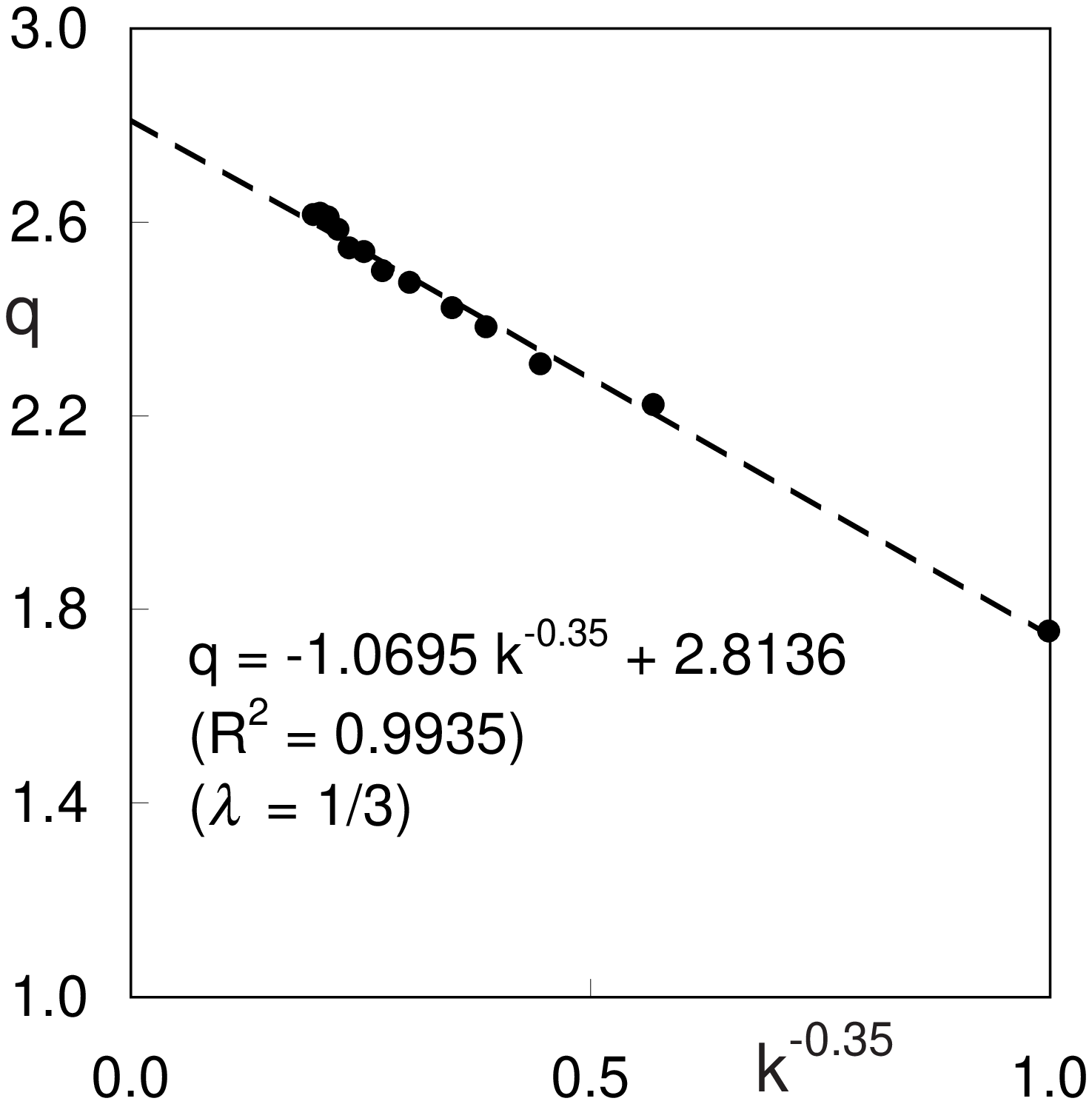}}}
\caption{\small Values of $q_{eq}$ obtained by Soares et al through the analysis of distributions emerging in the analysis of saddle point dynamics of the Henon-Heiles system. We have chosen the abscissa in such a way as to produce a linear form. $R^2$ is the square linear correlation factor. For the value $\lambda=1/3$ see [44].}
\end{figure}


\begin{thebibliography}{10}

\bibitem{tsallisjsp}C. Tsallis, J. Stat. Phys. {\bf 52}, 479 (1988); the bibliography of the subject is regularly updated at http://tsallis.cat.cbpf.br/biblio.htm
\bibitem{curado}E.M.F. Curado and C. Tsallis, J. Phys. A {\bf 24}, L69 (1991) [Corrigenda: {\bf 24}, 3187 (1991) and {\bf 25}, 1019 (1992)].
\bibitem{tsamepla}C. Tsallis, R.S. Mendes and A.R. Plastino, Physica A {\bf 261}, 534 (1998).
\bibitem{tsallisbjp}C. Tsallis, in {\it Nonextensive Statistical Mechanics and Thermodynamics}, eds. S.R.A. Salinas and C. Tsallis, Braz. J. Phys. {\bf 29}, 1 (1999) [accessible at http://www.sbf.if.usp.br/bjp/Vol29/Num1/index.htm].
\bibitem{tsallisspringer}C. Tsallis, in {\it Nonextensive Statistical Mechanics and its Applications}, eds. S. Abe and Y. Okamoto, Series {\it Lecture Notes in Physics} (Springer-Verlag, Berlin, 2001).
\bibitem{tsallisdenton}C. Tsallis, in {\it Classical and Quantum Complexity and Nonextensive Thermodynamics}, eds. P. Grigolini, C. Tsallis and B.J. West, Chaos , Solitons and Fractals {\bf 13}, Number 3, 371 (Pergamon-Elsevier, Amsterdam, 2002).
Solitons and Fractals {\bf 13}, Number 2 (Nov/Dec 2001), in press [cond-mat/0010150].
\bibitem{einstein}A. Einstein, B. Podolsky and N. Rosen, Phys. Rev. {\bf 47}, 777 (1935).
\bibitem{cohentannoudji}If we reintroduce in the formalism the Boltzmann constant $k_B$ with its nominal value in standard units, at the place of $(1-q)/T_q^{\prime}$ appears  $(1-q)/k_B T_q^{\prime}$ (or, more precisely, $(1-q)/k T_q^{\prime}$ with $k=f(q)k_B$, where $f(q)$ is some dimensionless function such that $f(1)=1)$. Therefore, the confluence onto BG statistics emerges for asymptotically large values of $(1-q)/k_B T_q^{\prime}$ , or equivalently in the limit $1/k_B \rightarrow 0$. This limit consistently re-emerges in the pseudoadditivity property of the entropic form $S_q$, which in usual units reads $S_q(A+B) = S_q(A)+S_q(B)+(1-q)S_q(A)S_q(B)/k_B$. Once again, in the limit $1/k_B \rightarrow 0$, the BG entropy additivity is satisfied. All this makes one to think about some analogies concerning mechanics. Indeed, in the limits $h \rightarrow 0$, $1/c \rightarrow 0$ and $G \rightarrow 0$, Newtonian mechanics emerges from quantum mechanics, special and general relativity. From this standpoint, we certainly agree with Cohen-Tannoudji's understanding \cite{cohentannoudji2} that the basic universal constants of the physical world (or at least of our contemporary understanding of it) are {\it four} in number ($c$, $h$, $G$ and $k_B$, referred to as Einstein, Planck, Newton and Boltzmann constants respectively) , and not only {\it three} ($c$, $h$ and $G$) as sometimes claimed. The subtle interplay of the universal constants can be illustrated, for instance, on Bose-Einstein and Fermi-Dirac statistics: They both converge onto Maxwell-Boltzmann statistics in the limit $h/k_B \rightarrow 0$, which can be indistinctively realized through $h \rightarrow 0$ or $1/k_B \rightarrow 0$. Along related lines, see also \cite{tsallisabe}.
\bibitem{cohentannoudji2}G. Cohen-Tannoudji, {\it  Les constantes universelles et la physique de l'horizon}, La Recherche {\bf 278}, 756 (1995)  (see also G. Cohen-Tannoudji, {\it Les constantes universelles}, Hachette, Paris, 1998).
\bibitem{tsallisabe}C. Tsallis and S. Abe, Physics Today  {\bf 51}, 114 (1998).
\bibitem{maxwell}J.C. Maxwell, Philos. Mag. (Ser. 4) 19 (1860) 19.
\bibitem{boltzmann}L. Boltzmann, Wien, Ber. {\bf 66}, 275 (1872); see also {\it Vorlesungen uber Gastheorie} (Leipzig, 1896) [{\it Lectures on Gas Theory}, transl. S. Brush (Univ. California Press, Berkeley, 1964].
\bibitem{gibbs}J.W. Gibbs, {\it Elementary Principles in Statistical Mechanics} (C. Scribner's Sons, New York, 1902; Yale University Press, New Haven, 1948).
\bibitem{darwinfowler}C.G. Darwin and R.H. Fowler, Phil. Mag. and J.Sci. {\bf 44}, 450 (1922); R.H. Fowler, Phil. Mag. and J. Sci. {\bf 45}, 497 (1923).
\bibitem{khinchin1}A.I. Khinchin, {\it Mathematical Foundations of Statistical Mechanics} (Dover, 1949). 
\bibitem{balian}R. Balian and N.L. Balazs, Ann. Phys. (NY) {\bf 179}, 97 (1987).
\bibitem{kubo}R. Kubo, H. Ichimura, T. Usui and N. Hashitsume, {\it Statistical Mechanics} (North-Holland, Amsterdam, 1988).
\bibitem{shannon}C.E. Shannon, Bell System Tech. J. {27}, 379 and 623 (1948).
\bibitem{khinchin2}A.I. Khinchin, Uspekhi Matem. Nauk {\bf 8}, 3 (1953) [transl. R.A. Silverman and M.D. Friedman, {\it Mathematical Foundations of Information Theory} (Dover, New York, 1957)].
\bibitem{ardarwinfowler}S. Abe and A.K. Rajagopal, J. Phys. A {\bf 33}, 8733 (2000).
\bibitem{arkhinchin}S. Abe and A.K. Rajagopal, Europhys. Lett. {\bf 52}, 610 (2000).
\bibitem{arbalian}S. Abe and A.K. Rajagopal, Phys. Lett. A {\bf 272}, 341 (2000).   
\bibitem{arkubo}S. Abe and A.K. Rajagopal, Europhys. Lett. {\bf 55}, 6 (2001). 
\bibitem{lima}J.A.S. Lima, R. Silva and A.R. Plastino, Phys. Rev. Lett. {\bf 86}, 2938 (2001).  
\bibitem{kaniadakis}G. Kaniadakis, Physica A {\bf 296}, 405 (2001).
\bibitem{santos}R.J.V. Santos, J. Math. Phys. 38, 4104 (1997).
\bibitem{abekhinchin}S. Abe, Phys. Lett. A {\bf 271}, 74 (2000).
\bibitem{levy}D.H. Zanette and P.A. Alemany, Phys. Rev. Lett. {\bf 75}, 366 (1995); C. Tsallis, S.V.F Levy, A.M.C. de Souza and R. Maynard, Phys. Rev. Lett. {\bf 75}, 3589 (1995) [Erratum: {\bf 77}, 5442 (1996)]; M.O. Caceres and C.E. Budde, Phys. Rev. Lett. {\bf 77}, 2589 (1996); D.H. Zanette and P.A. Alemany, Phys. Rev. Lett. {\bf 77}, 2590 (1996); D. Prato and C. Tsallis, Phys. Rev. E {\bf 60}, 2398 (1999); A. Robledo, Phys. Rev. Lett. {\bf 83}, 2289  (1999).
\bibitem{correlated}A.R. Plastino and A. Plastino, Physica A  {\bf 222}, 347 (1995); C. Tsallis and D.J. Bukman, Phys. Rev. E {\bf 54}, R2197 (1996); A. Compte and D. Jou, J. Phys. A {\bf  29}, 4321 (1996).
\bibitem{galaxy}A. Lavagno, G. Kaniadakis, M. Rego-Monteiro, P. Quarati and C. Tsallis, Astrophysical Letters and Communications {\bf  35}, 449 (1998).
\bibitem{bogho}B.M. Boghosian, Phys. Rev. E {\bf 53}, 4754 (1996).
\bibitem{ramos}F.M. Ramos, R.R. Rosa and C. Rodrigues Neto, cond-mat/9907348; F.M. Ramos, C. Rodrigues Neto and R.R. Rosa, cond-mat/0010435.
\bibitem{turbulence} T. Arimitsu and N. Arimitsu, Phys. Rev. E {\bf 61}, 3237 (2000);  T. Arimitsu and N. Arimitsu, J. Phys. A {\bf 33}, L235 (2000).
\bibitem{turbulencebeck}C. Beck, Physica A {\bf 277}, 115 (2000).
\bibitem{swinney}C. Beck, G.S. Lewis and H.L. Swinney, Phys. Rev. E {\bf 63}, 035303 (2001).
\bibitem{citations}C. Tsallis and M.P. de Albuquerque, Eur. Phys. J. B {\bf 13}, 777 (2000).
\bibitem{montemurro}M.A. Montemurro, Physica A (2001), in press [cond-mat/0104066].
\bibitem{bemski}C. Tsallis, G. Bemski and R.S. Mendes, Phys. Lett. A {\bf 257}, 93 (1999).
\bibitem{rajagopal}S. Abe and A.K. Rajagopal, Phys. Rev. A {\bf 60}, 3461 (1999); A. Vidiella-Barranco, Phys. Lett. A {\bf 260}, 335 (1999).
\bibitem{sethmichel}C. Tsallis, S. Lloyd and M. Baranger, Phys. Rev. A {\bf 63}, 042104 (2001).; C. Tsallis, P.W. Lamberti and D. Prato, Physica A {\bf 295}, 158 (2001).
\bibitem{bediaga}I. Bediaga, E.M.F. Curado and J. Miranda, Physica A {\bf 286}, 156 (2000); C. Beck, Physica A {\bf 286}, 164 (2000).
\bibitem{rafelski}D.B. Walton and J. Rafelski, Phys. Rev. Lett. {\bf 84}, 31 (2000).
\bibitem{ivano}H.P. de Oliveira, S.L. Sautu, I.D. Soares and E.V. Tonini, Phys. Rev. D {\bf 60}, 121301-1 (1999).
\bibitem{oliveirasoares}H.P. de Oliveira, I.D. Soares and E.V. Tonini, Physica A {\bf 295}, 348 (2001).
\bibitem{ion}D.B. Ion and M.L.D. Ion, Phys. Rev. Lett. {\bf 81}, 5714 (1998); M.L.D. Ion and D.B. Ion, Phys. Rev. Lett. {\bf 83}, 463 (1999); D.B. Ion and M.L.D. Ion, Phys. Rev. E {\bf 60}, 5261 (1999); D.B. Ion and M.L.D. Ion,  in {\it Classical and Quantum Complexity and Nonextensive Thermodynamics}, eds. P. Grigolini, C. Tsallis and B.J. West, Chaos , Solitons and Fractals {\bf 13}, Number 3, 547 (Pergamon-Elsevier, Amsterdam, 2002).; M.L.D. Ion and D.B. Ion, Phys. Lett. B {\bf 474}, 395 (2000); M.L.D. Ion and D.B. Ion, Phys. Lett. B {\bf 482}, 57 (2000); D.B. Ion and M.L.D. Ion, Phys. Lett. B {\bf 503}, 263 (2001).
\bibitem{arpita}A. Upadhyaya, J.-P. Rieu, J.A. Glazier and Y. Sawada, Physica A {\bf 293}, 549 (2001).
\bibitem{maps}C. Tsallis, A.R. Plastino and W.-M. Zheng, Chaos, Solitons and Fractals {\bf 8}, 885 (1997); U.M.S. Costa, M.L. Lyra, A.R. Plastino and C. Tsallis, Phys. Rev. E {\bf 56}, 245 (1997); M.L. Lyra and C. Tsallis, Phys. Rev. Lett. {\bf 80}, 53 (1998); M.L. Lyra, Ann. Rev. Comp. Phys. , ed. D. Stauffer (World Scientific, Singapore, 1998), page 31; V. Latora, M. Baranger, A. Rapisarda and C. Tsallis, Phys. Lett. A {\bf 273}, 97 (2000); U. Tirnakli, G.F.J. Ananos and C. Tsallis, Phys. Lett. A (2001), in press [cond-mat/0005210]. 
\bibitem{moura}F.A.B.F. de Moura, U. Tirnakli and M.L. Lyra, Phys. Rev. E {\bf 62}, 6361 (2000).
\bibitem{cataniario} C. Anteneodo and C. Tsallis, Phys. Rev. Lett. {\bf 80}, 5313 (1998); A. Campa, A. Giansanti, D. Moroni and C. Tsallis, Phys. Lett. A {\bf 286}, 251 (2001).
\bibitem{cataniario2}V. Latora, A. Rapisarda and C. Tsallis, {\it Non-Gaussian equilibrium in a long-range Hamiltonian system}, Phys. Rev. E (2001), in press [cond-mat/0103540]; V. Latora, A. Rapisarda and C. Tsallis, {\it Fingerprints of nonextensive thermodynamics in a long-range hamiltonian system}, cond-mat/0109056; A. Campa, A. Giansanti and D. Moroni, {\it Metastable states in a class of long-range Hamiltonian systems}, cond-mat/0109178.
\bibitem{ernestogarin}E. P. Borges, C. Tsallis, G. Ananos and P.M.C. Oliveira, to be published.
\bibitem{baldovin}F. Baldovin, C. Tsallis and B. Schulze, {\it  Nonstandard mixing in the standard map}, cond-mat/0108501; F. Baldovin, {\it Mixing and approach to equilibrium in the standard map}, cond-mat/0109356.
\bibitem{fulviocasati}F. Baldovin, C. Tsallis and G. Casati, (2001) in progress.
\bibitem{krylov}N. Krylov, Nature {\bf 153}, 709 (1944); N.S. Krylov, {\it Works on the 
Foundations of Statistical Physics}, translated by A.B. Migdal, Ya. 
G. Sinai and Yu. L. Zeeman, Princeton Series in Physics (Princeton University Press, Princeton, 1979).
\bibitem{becklagrangian}C. Beck, Phys. Lett. A {\bf 287}, 240 (2001).
\bibitem{lubbe}J.C.A. van der Lubbe, {\it Information Theory} (Cambridge University Press, Cambridge, 1997), p. 171.
\bibitem{cover}T.M. Cover and J.A. Thomas, {\it Elements of Information Theory} (J. Wiley and Sons, New York, 1991), Chapter 9.

\end{thebibliography}
\end{document}